\documentclass[useAMS,usenatbib]{mn2e}
\usepackage{graphicx,rotating,subfigure,floatflt,amsmath,wasysym}
\usepackage{graphicx,subfigure,floatflt,amsmath,wasysym, amssymb}
\usepackage{here}
\usepackage{setspace}
\usepackage{lscape}

\title[Polar-ring galaxy in the Subaru Deep Field]{A candidate polar-ring galaxy in the Subaru Deep Field}
\author[Ido Finkelman, Or Graur \& Noah Brosch]{Ido Finkelman\thanks{E-mail:
ido@wise.tau.ac.il}, Or Graur \& Noah Brosch\\
The Wise Observatory and the Raymond and Beverly Sackler School of Physics and
Astronomy, the Faculty of Exact Sciences, \\ Tel Aviv University, Tel Aviv 69978, Israel}
\begin{document}

\date{Accepted 2010 October 20. Received 2010 October 14; in original form 2010 August 24}

\pagerange{\pageref{firstpage}--\pageref{lastpage}} \pubyear{2002}

\maketitle

\label{firstpage}

\begin{abstract}
We discuss the properties of an object in the Subaru Deep Field (SDF) classified as a galaxy in on-line data bases and revealed on the Subaru images as a genuine polar-ring galaxy (PRG) candidate.
We analyse available photometric data and conclude that this object consists of a $\gtrsim5$ Gyr old early-type central body surrounded by a faint, narrow inner ring tilted at a $\sim25^\circ$ angle relative to the polar axis of the host galaxy. The halo surrounding the main stellar body exhibits a diversity of spatially extended stellar features of low surface brightness, including a faint asymmetric stellar cloud and two prominent loops. These faint features, together with the unperturbed morphology of the central host, are clear signs of a recent coalescence of two highly unequal mass galaxies, most likely a pre-existing early-type galaxy and a close-by gas-rich dwarf galaxy. 
The presumed stellar remnants observed near the edges of the ring, including possibly the surviving captured companion itself, indicate that the merger is still taking place. 

\end{abstract}

\begin{keywords}
galaxies: peculiar - galaxies: individual: SDSS J132533.22+272246.7 - galaxies: photometry
\end{keywords}

\section{Introduction}
Mutual interactions between galaxies are probably one of the main processes leading to
the formation of presently observed galaxies. 
The discovery of shells and ripples around elliptical galaxies (Kormendy \& Djorgovski 1989; Tal et al.\ 2009) and of extended stellar tidal features around normal disc galaxies (Mart\'{i}nez-Delgado et al.\ 2010) are among the most compelling evidence for this scenario.
In more extreme and rare cases the remnants of the interacting galaxies are not mixed in one smooth object, but some of the
material settles quasi-statically into the equatorial plane of the main body, forming a polar-ring galaxy (PRG; Whitmore et al.\ 1990). 
Alternatively, simulations show that the polar structure can form as the result of a ``secondary event'' during the evolution of the galaxy, involving the accretion of gas from a close, gas-rich dwarf companion (Reshetnikov \& Sotnikova 1997). 

However, both formation scenarios are difficult to validate observationally since they predict the formation of a polar ring on short time scales of less than 1-2 Gyr (Bekki 1998; Bournaud \& Combes 2003). Since only about 0.5 per cent of all early-type galaxies have observable polar rings (Whitmore et al.\ 1990), on-going galaxy interactions that form a PRG are probably extremely rare events. 
A galaxy tidally disrupted while still in orbit around a much more luminous and massive companion should leave distinctive tidal stellar features before merging with the central galaxy or mixing into a seemingly smooth component. These stellar debris can extend well beyond the central galaxy and mostly occur at a very low surface brightness level, making it difficult to detect in sky survey images. 

In this paper we present an interpretation of existing observational data of the object SDSS J132533.22+272246.7 (hereafter referred to as RG)  identified as a ring galaxy during a supernova survey in the Subaru Deep Field (SDF; Graur et al., in preparation). The object is centred at $\alpha=13^{h}25^{m}33.2^{s}, \delta=+27^{\circ}22\arcmin46.7\arcsec$ (J2000) at a redshift of $z=0.061$, which puts it at a distance of about 250 Mpc, assuming $H_0=73$ km sec$^{-1}$ Mpc$^{-1}$. This paper is organized as follows: Section \ref{obs} gives a description of the observations and data reduction; in Section \ref{results} we present the observational results along with our interpretation of the object and the main conclusions are summarized in Section \ref{conc}. 

\begin{figure*}
  \caption{Panel a: SDSS $g$-band image of RG. Panels b-d: Combined Subaru image displayed in different contrast levels to enhance structural features. Stellar debris and loops are marked by arrows. \label{RG}}
  \vspace{2mm}
\begin{center}
\begin{tabular}{cc}
\includegraphics[trim=0.5cm 0cm 0.5cm 0.5cm,width=6cm]{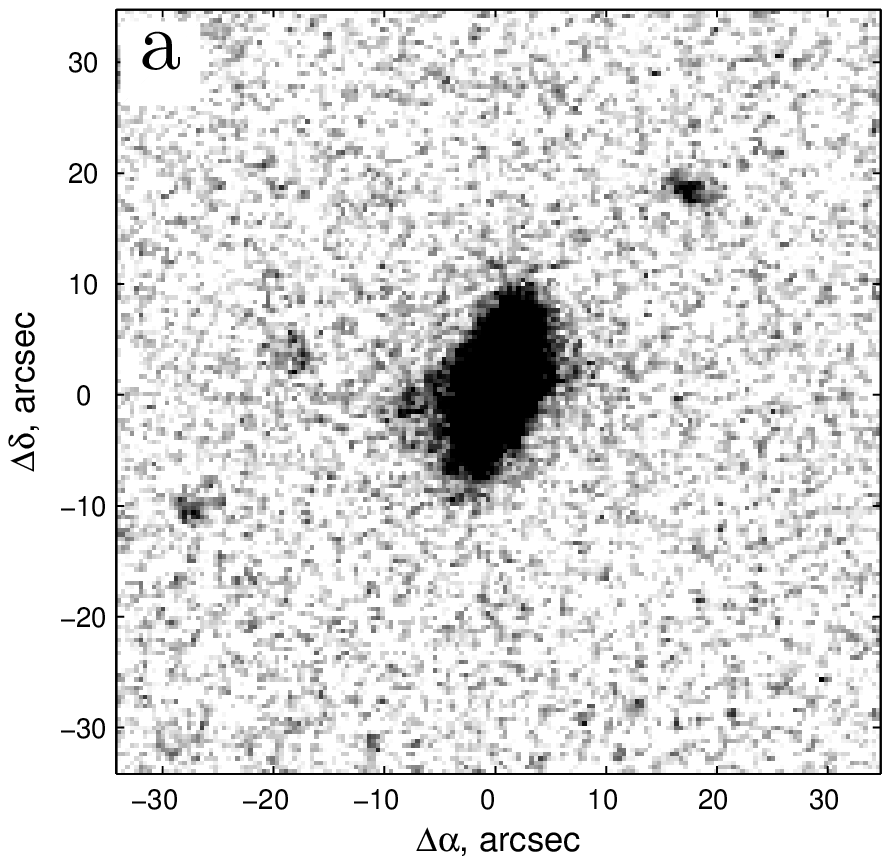} & \includegraphics[trim=0.5cm 0cm 0.5cm 0.5cm,width=6cm]{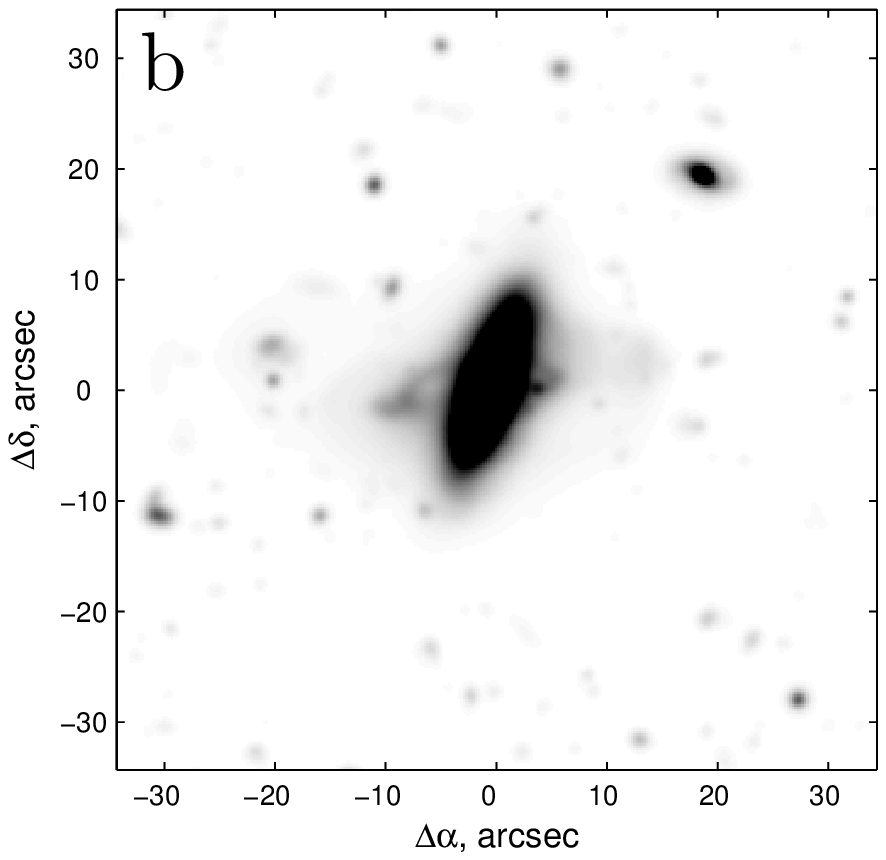} \\
\vspace{5cm}
\includegraphics[trim=0.5cm 0cm 0.5cm 0.5cm,width=6cm]{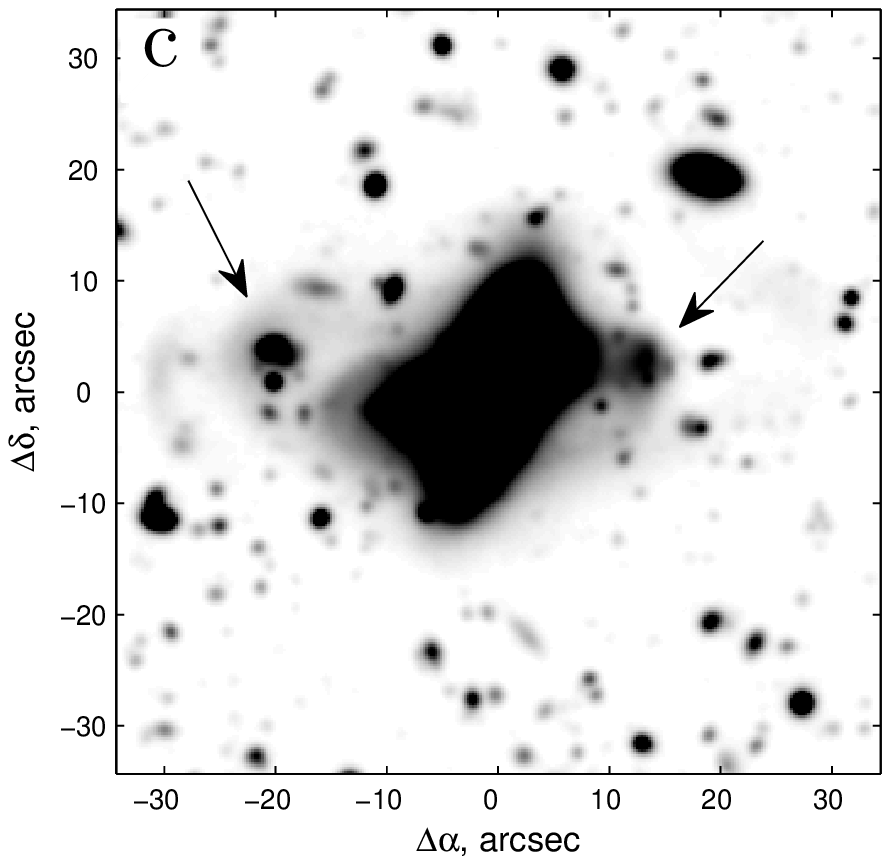} & \includegraphics[trim=0.5cm 0cm 0.5cm 0.5cm,width=6cm]{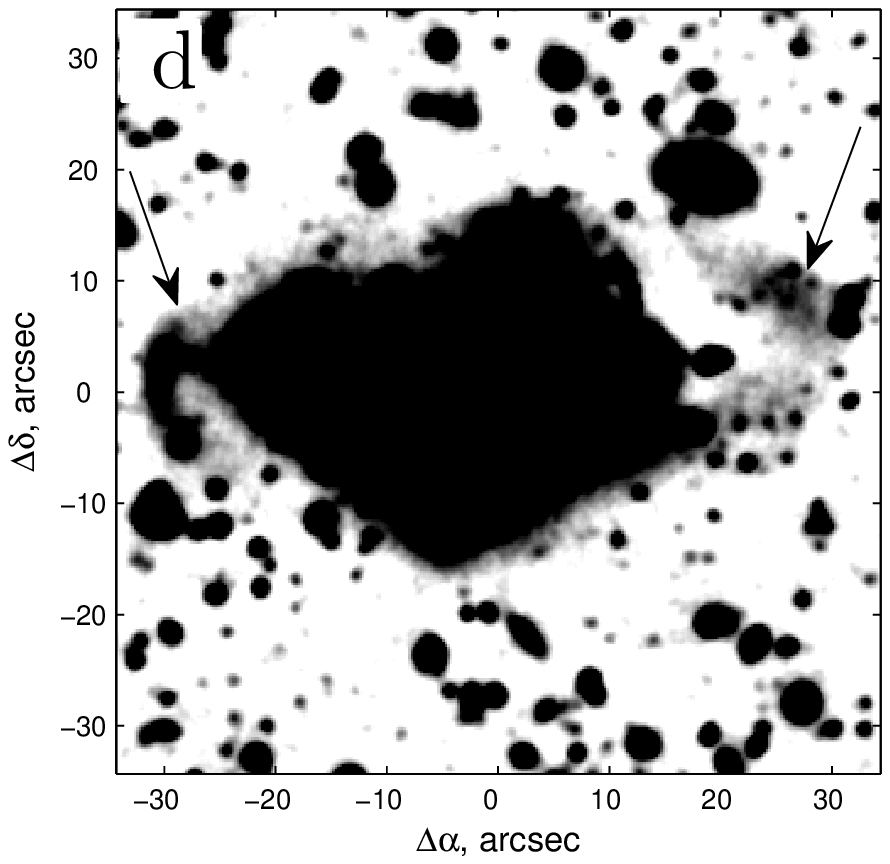} 
\vspace{-5cm}
\end{tabular}
\end{center}
\end{figure*}

\section{Observations and data reduction}
\label{obs}
Here we make use of available archived Subaru data of the SDF, a 0.25 deg$^2$ field relatively devoid of bright stars or large galaxies.
The SDF was imaged by Kashikawa et al.\ (2004) with the Suprime-Cam camera on the Subaru 8.2-m telescope situated on Mauna Kea, Hawaii. 
The observations included five broad-band filters ({\it B}, {\it V}, {\it R}, {\it i'} and {\it z'}) with an average seeing of $1$ arcsec in all bands.
Individual exposures were reduced with the Suprime-Cam pipeline SDFRED (Yagi 2002; Ouchi et al.\ 2004) including overscan subtraction, flat fielding, distortion correction and sky subtraction. The images were then registered and combined into separate bands. 
The Subaru data are calibrated to magnitudes in the AB system.
We also obtained publicly available processed and photometrically calibrated data of RG from the Sloan Digital Sky Survey (SDSS), the Two Micron All Sky Survey (2MASS) and Galaxy Evolution Explorer (GALEX). 
\section{Results and discussion}
\label{results}
RG appears on the SDSS images as an early-type galaxy surrounded by a faint, elusive ring which does not extend farther than the diameter of the host (see {\it g}-band image in Fig.\ \ref{RG}a). 
The SDSS database provides photometry of the central object ($m_{\it g}=16.70$) and spectroscopic data of the central $3$ arcsec ($\simeq3.6$ kpc) region of the galaxy. 
A visual inspection of the Subaru images clearly reveals the presence of an inclined tight faint ring around the central body.
To emphasize the faint outer structure we present in Fig.\ \ref{contour} a {\it B}-band contour map of the object with isophotes drawn at 1 mag arcsec$^{-2}$ intervals from 21 to 31 mag arcsec$^{-2}$. 
An even deeper image of the galaxy is produced by combining the different broad-band images. The combined image, displayed at different contrast levels in Figs \ref{RG}b-d, has an effectively wider photometric band which is only useful to examine the morphology and the fainter and outer regions of the galaxy. These figures clearly indicate that the host galaxy contributes most of the RG light.

\subsection{Host galaxy}
\label{S:host}
\begin{figure}
  \caption{{\it B}-band contour map. Contour levels are from 21 to 31 mag arcsec$^{-2}$ in 1 mag arcsec$^{-2}$ steps. \label{contour}}
  \centering
 \vspace{0.2cm}
 \includegraphics[trim=0cm 0cm 0.5cm 0.5cm,width=8cm]{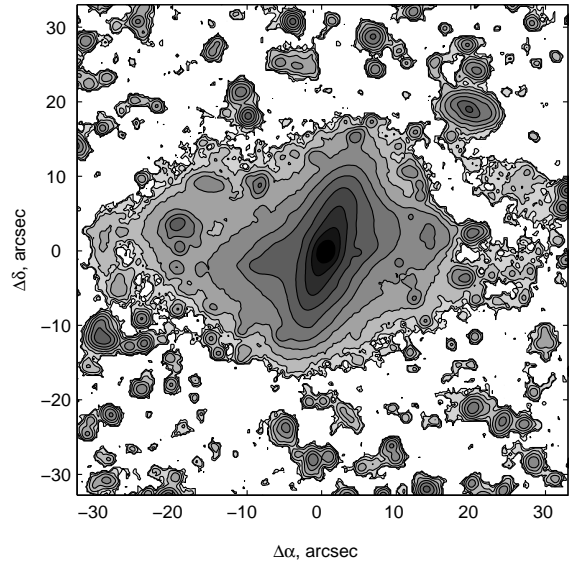} 
\end{figure}

Since the isophotes of RG appear almost identical in all bands, we make use of the {\it B}-band isophotes to discuss the structural morphology of the host galaxy.
The $\mu_B=25$ mag arcsec$^{-2}$ isophote lies at a semi-major axis distance of $10.3$ arcsec. 
Measuring the intensity along isophotes using the ELLIPSE task in {\small IRAF}\footnote{{\small IRAF} is distributed by the National Optical Astronomy Observatories (NOAO), which is operated by the Association of Universities, Inc. (AURA) under co-operative agreement with the National Science Foundation} we find the position angle to be constant at $\sim161.5^\circ$ outwards from the centre up to $\mu_B\simeq24.0$ mag arcsec$^{-2}$, whereas the ellipticity increases from $\sim0.2$ to $\sim0.6$. This indicates that a bulge dominates the inner regions while the outer isophotes are more flattend, probably due to the presence of a discy structure, as typical of an early-type galaxy (see also Fig.\ \ref{contour}).

To further investigate the nature of the host we used the high-pass median filtering (unsharp masking) technique to enhance fine structures on the Subaru images. The right panel in Fig.\ \ref{moreRG} presents the residual {\it R}-band image, obtained by subtracting the median filtered image from the original image. The resultant image reveals a highly inclined disc component lying along the major-axis of the host galaxy. The {\it B}-{\it R} colour map of RG, shown in the left panel of Fig.\ \ref{moreRG}, further supports this conclusion. The disc shows a red gradient decreasing from east to west, presumably due to internal extinction.

A two-component model consisting of a S\'{e}rsic bulge (an $r^{1/n}$ component) and an exponential disc was fitted to the surface brightness profile along the major-axis of the host, assuming it is not affected by the light coming from the ring. 
Since the central pixels of RG in the  Subaru images saturate, we used the SDSS {\it g}, {\it r} and {\it i}-band images to compute the surface brightness profile in the inner parts of the galaxy. 
Our least-squares algorithm yielded the best-fit parameters for the 1D model: a S\'{e}rsic index, effective radius and effective surface magnitude of $n\simeq1.4$, $r_e\simeq0.9$ arcsec and $\mu_{b,e}\simeq20.0$ mag arcsec$^{-2}$ for the bulge, and a scale length and central surface brightness magnitude of $h\simeq2.9$ arcsec and $\mu_{d,0}\simeq20.8$ mag arcsec$^{-2}$ for the disc (see Iodice et al.\ 2002a). Taking into account the inclination of the disc, we estimate a total bulge-to-disc ratio of $\sim1.1$, typical of an S0 or Sa galaxy (Kent 1985). The major-axis light profile is plotted along with the fitted model in Fig.\ \ref{profiles}. The model matches well the observed light profile in the inner regions where the light from the central body dominates, whereas larger deviations are observed in the outer regions.

We also obtained integrated colours from the SDSS, 2MASS and GALEX databases and used them to estimate the host galaxy age. 
This was done by comparing the observed colours with stellar population synthesis models developed by Bruzual \& Charlot (2003) for different star formation histories (see Iodice et al.\ 2002b). 
We found that the host colours are well-reproduced by an old stellar population formed in an instantaneous burst at least 5 Gyr ago. We note that the colours are redder in the inner 3 arcsec region of the host and that stellar absorption features dominate the SDSS spectra of RG with no significant line emission. 
The integrated magnitudes and colours of the host galaxy are listed in Table \ref{RGtable}.
\subsection{Ring}
\begin{figure*}
  \caption{Left panel: {\it B}-{\it R} colour index map of RG. The grey scale is such that bluer regions appear darker. Right Panel: The {\it R}-band residual image obtained by subtracting the median filtered image. Note the saturated pixels in the inner part of the galaxy. \label{moreRG}}
\vspace{2mm}
\begin{center}
\begin{tabular}{cc}
\includegraphics[trim=0.5cm 0.5cm 0cm 0.5cm,width=6cm]{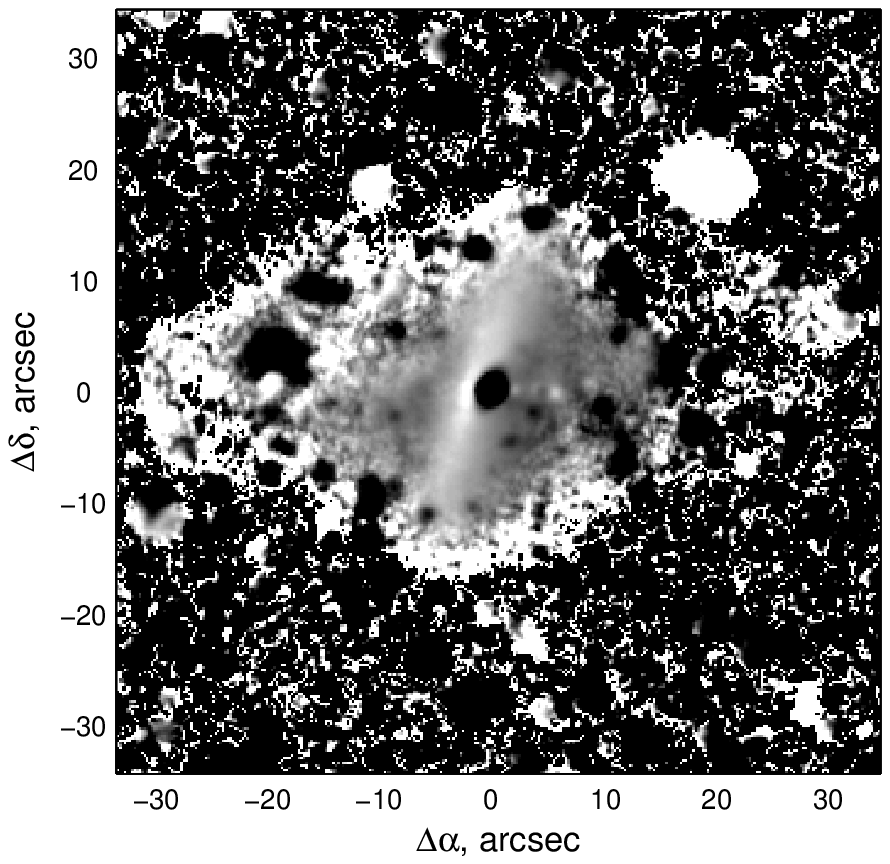} & \includegraphics[trim=0cm 0.5cm 0.5cm 0.5cm,width=6cm]{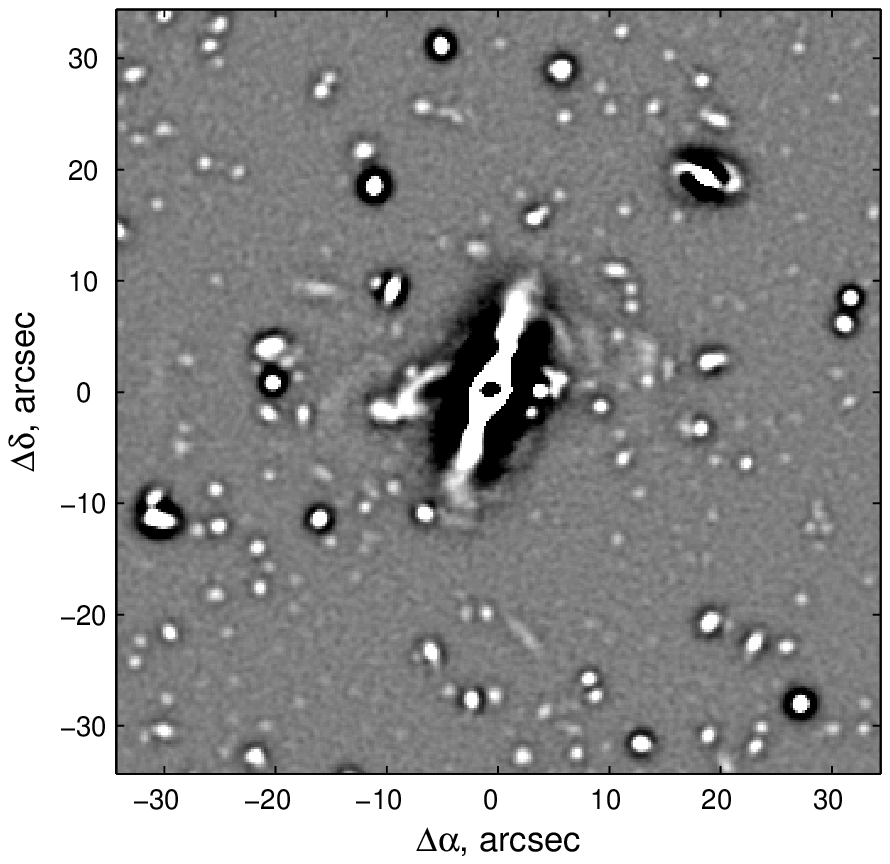} 
\end{tabular}
\end{center}
\label{fig:}
\end{figure*}

\begin{figure*}
  \caption{Upper panels represent the surface brightness profiles obtained along the major-axis of the host galaxy (left) and the long axis of the ring (right). The solid line represents the 1D two-component model fit to the observations. Lower panels represent colour variation along the axes. The surface brightness values within the central 5 arcsec were measured from the SDSS images. Although to smooth the profiles the values are plotted in one-pixel steps, the data has no physical meaning on scales smaller than $\sim1.2$ arcsec (the typical FWHM).  \label{profiles}}
\vspace{2mm}
\begin{center}
\begin{tabular}{cc}
\includegraphics[trim=0.5cm 0.5cm 0cm 0.5cm,width=6cm]{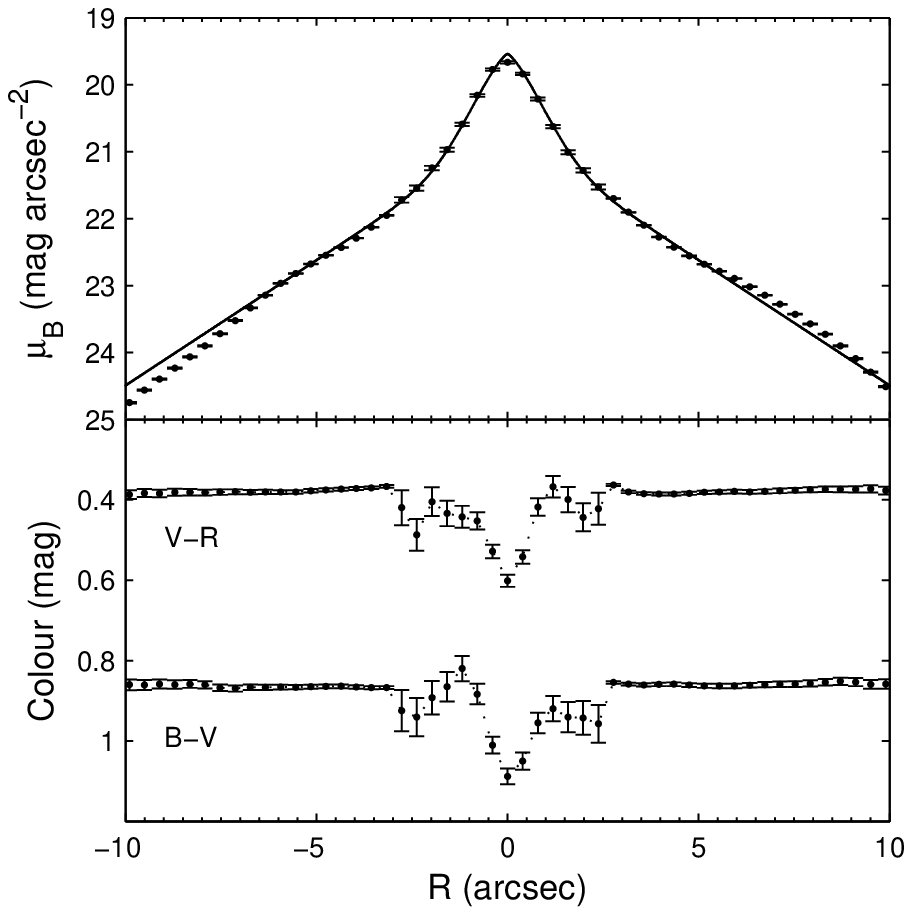} & \includegraphics[trim=0cm 0.5cm 0.5cm 0.5cm,width=6cm]{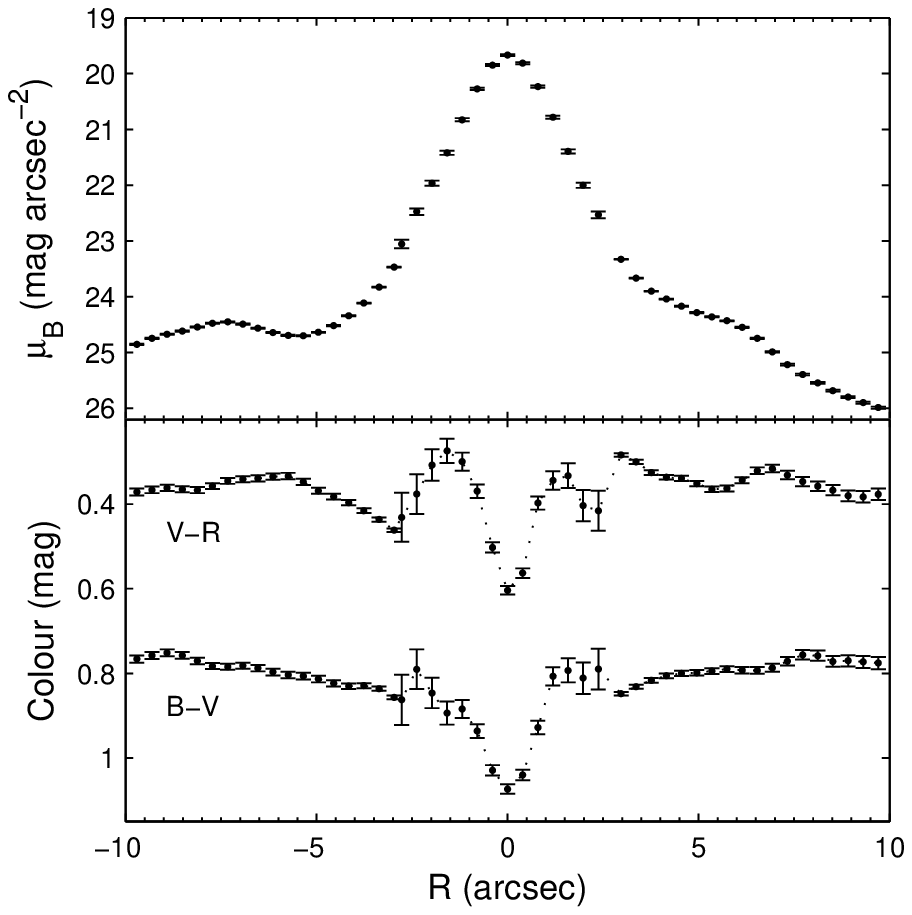}
\end{tabular}
\end{center}
\label{fig:}
\end{figure*}

The tight and curved edges of the bright part of the ring are clearly seen in the residual image presented in Fig.\ \ref{moreRG}, implying that the off-centre ring is $\sim2\pm1$ arcsec in width. The surface brightness along the apparent long axis of the ring is presented in Fig.\ \ref{profiles}. The ring is tilted by about $25^\circ$ to the minor axis of the host and by about $70^\circ$ to the plane of the sky, assuming it is circular. 
The ring is clearly asymmetric, as shown by the light profile, and its surface brightness slightly increases to a maximum of $\sim24.5$ mag arcsec$^{-2}$ at 7.5 arcsec on the east side, while on the west side it exhibits a shoulder at 6 arcsec. 

A more detailed analysis of RG, including its appearance at different contrast levels, reveals an extended low surface brightness envelope around the ring. On its west side, the faint material connects with a fragmented, non-stellar object which presumably consists of stars recently captured by RG during an interaction with a passing companion. On the opposite side, at an intermediate angle between the ring and the host, we detected a brighter disrupted object which is very likely the surviving companion (see Fig.\ \ref{RG}c). Further out, at $\sim25-30$ arcsec on both sides of the central galaxy, are also two extensive circular features of very low surface brightness ($\mu_B\gtrsim28.0$ mag arcsec$^{-2}$; see Figs \ref{RG}d and \ref{contour}). 
Deep spectroscopic observations are necessary to confirm that the presumed stellar debris are physically connected with RG.

\begin{figure}
  \caption{Contour plots delineating the different areas where the integrated magnitudes are computed. The elliptical contour at the centre represents an extrapolation of the $\mu_{B}=27$ mag arcsec$^{-2}$ isophote of the host galaxy based on the morphological analysis in Section \ref{S:host}. The two lobes on each side of the host enclose the area where the surrounding stellar halo, including the ring, is brighter than $\mu_{B}=27$ mag arcsec$^{-2}$. \label{areas}}
\begin{center}
\includegraphics[trim=0.5cm 0.5cm 0.5cm 0cm,width=6cm]{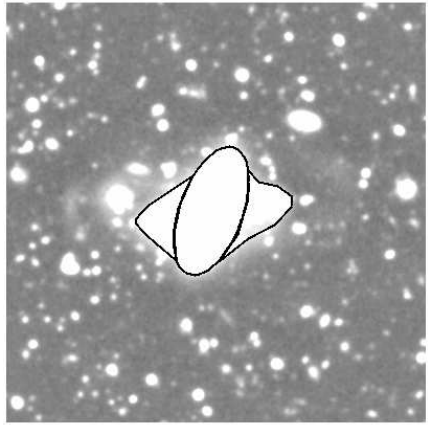} 
\end{center}
\label{fig:}
\end{figure}

We measure the integrated magnitudes and colours of the stellar halo around the ring in parts where it is less contaminated by the host galaxy, as illustrated in Fig.\ \ref{areas}, and list them in Table \ref{RGtable}. Our results clearly indicate that the light coming from the ring and the halo constitutes only a small fraction of the total light of the object. The surrounding stellar halo, including the ring, is bluer than the host galaxy, but is not nearly as blue as reported for other PRGs (e.g., NGC 4650A).
\begin{table*}
  \caption{Photometric properties of the regions presented in Fig.\ \ref{areas}. Magnitudes are not K-corrected. \label{RGtable}}
\begin{tabular}{llrrrr}
\hline
 Photometry & Source & SDSS J132533.22+272246.7 & Host galaxy & Stellar halo \\
 \hline
 m$_{\it g}$ & SDSS & $16.70\pm0.01$ & - & - \\
 m$_{\it B}$ & Subaru & $17.10\pm0.03$  & $17.16\pm0.03$ & $20.38\pm0.01$ \\
 M$_{\it B}$ & Subaru & $-19.97\pm0.03$ & $-19.91\pm0.03$ & $-16.69\pm0.01$\\
 {\it B}-{\it V} & Subaru & - & $0.85\pm0.04$ & $0.75\pm0.02$ \\
 {\it V}-{\it R} & Subaru & - & $0.37\pm0.04$ & $0.39\pm0.02$\\
 {\it R}-{\it i'} & Subaru & - & $0.24\pm0.04$ & $0.13\pm0.02$\\
 {\it i'}-{\it z'} & Subaru & - & $0.32\pm0.04$ & $0.12\pm0.02$\\
 m$_{\it NUV}$ & GALEX & $21.38\pm0.01$ & - & -\\
 {\it FUV}-{\it NUV} & GALEX & $1.82\pm0.08$ & - & -\\
 m$_{\it J}$ & 2MASS & $14.22\pm0.05$ & - & -\\
 {\it J}-{\it H} & 2MASS & $0.70\pm0.08$ & - & -\\ 
 {\it H}-{\it K} & 2MASS & $0.35\pm0.09$ & - & -\\
 \hline
 \end{tabular}
\end{table*}
\subsection{PRG formation}
PRGs are believed to form as a result of a dissipative galaxy merger between two disc galaxies, or by the accretion of gas from a small gas-rich companion by a pre-existing S0 galaxy (see Bekki 1998; Bournaud \& Combes 2003 and references therein). 
Distinguishing between these two kinds of scenarios can be addressed by examining their morphology and by observing polar-ring formation in action in the local universe.

PRGs are typically divided into two groups: (1) galaxies with an unusually compact and faint bulge and an extended disc-like ring with the central region cut out; (2) galaxies with a dominant bulge and a narrow ring which is not extended in radius (Whitmore 1991; Reshetnikov \& Sotnikova 1997). RG seems to fit into the second category. Among known PRGs, RG resembles ESO 415-G26 by its faint narrow ring (see e.g.\ fig.\ 2 of Iodice et al.\ 2002a), central body morphology, red colours and the presence of tidal features extending beyond the host (e.g.\ loops and debris, see Whitmore et al.\ 1987).
 
The morphology of a PRG formed in a merger event depends not only on the orbital parameters, but also on the initial mass ratio of the two disc galaxies. For instance, a mass ratio larger than $\sim1.5$ can reproduce both the central discy S0 host and the inclined narrow ring of ESO 415-G26 (Bekki 1998). In this view, the diffusely dispersed material around RG can be explained as stars pushed out from the victim galaxy during a head-on collision with a larger, more massive galaxy.
The faint stellar features around RG are due to polar material oscillating at large and small radii before gathering in the ring, and the interacting companion is expected to eventually merge with the central object (Bournaud \& Combes 2003).
This scenario, previously proposed by Whitmore, McElroy \& Schweizer (1987) for ESO 415-G26, places strong constraints on the age of the system. 
Considering the off-centre position of the ring and the distinctive structural features surrounding the main body, we conclude that RG is only a few dynamical periods old and is still developing (see also Schweizer, Whitmore \& Rubin  1983).
  
A scenario in which gas was accreted from a gas-rich companion onto RG can be safely ruled out. 
In such a scenario, no clear sign relating the ring and the accreted companion is expected to be left even 1 or 2 Gyr after the ring has formed. The accreted matter, including only few stars, falls directly into a polar orbit and therefore a stellar 
halo around the polar ring is not expected, contrary to the merging scenario (Bournaud \& Combes 2003).

Understanding the role played by different mechanisms in PRG formation requires a deep photometric study of a large sample of such objects at various redshifts.
The redshift of RG places it between the high-$z$ PRG-candidates found in {\it HST} observations (Reshetnikov 1997; Lavery et al.\ 2004; Reshetnikov \& Dettmar 2007) and the low-$z$ PRGs published and studied by Whitmore et al.\ (1990) and others. 
However, observed PRGs that could bridge the distance gap are very rare since they are difficult to detect by large and shallow surveys such as the SDSS (see Brosch et al.\ 2010). In particular, detecting tidal features or merger remnants in such systems often requires exquisite surface brightness sensitivity (Tal et al.\ 2009; Smirnova, Moiseev \& Afanasiev 2010; Mart\'{i}nez-Delgado et al.\ 2010). RG is an excellent example, since the polar ring and the complex stellar halo structure could probably not have been detected by a cursory examination of sky survey images. 

\section{Conclusions}
\label{conc}
We present a photometric study of a candidate PRG in the SDF. 
The object has a complex structure consisting of:\\
1. A central host which we identify as a $\gtrsim5$ Gyr old early-type galaxy with an exponential bulge and a nearly edge-on disc;\\
2. A faint off-centre narrow ring at a $\sim25^\circ$ angle to the polar axis of the host; \\
3. A low surface brightness stellar halo extending asymmetrically on both sides of the ring;\\
4. At least two disrupted stellar debris in intermediate position angles between the ring and the polar axis of the host;\\
5. Two outer loops on both sides in the direction of the stellar debris.\\
Our morphological study suggests that this galaxy belongs to the group of bulge-dominated PRGs and that the ring was probably formed around the pre-existing host during a merger with a gas-rich dwarf companion. This event probably took place less than 1-2 Gyr ago, since it appears that the captured material has not yet fully dispersed into the halo or settled into the polar structure.    

\section*{Acknowledgments}
The authors wish to recognize and acknowledge the very significant cultural role and reverence that the summit of Mauna Kea has always had within the indigenous Hawaiian community. We are most fortunate to have the opportunity to conduct observations from this mountain.

Based on observations made with the NASA Galaxy Evolution Explorer.
GALEX is operated for NASA by the California Institute of Technology under NASA contract NAS5-98034.

This publication makes use of data products from the Two Micron All Sky Survey, which is a joint project of the University of Massachusetts and the Infrared Processing and Analysis Center/California Institute of Technology, funded by the National Aeronautics and Space Administration and the National Science Foundation.

Funding for the SDSS and SDSS-II has been provided by the Alfred P. Sloan Foundation, the Participating Institutions, the National Science Foundation, the U.S. Department of Energy, the National Aeronautics and Space Administration, the Japanese Monbukagakusho, the Max Planck Society, and the Higher Education Funding Council for England. The SDSS Web Site is http://www.sdss.org/.

The SDSS is managed by the Astrophysical Research Consortium for the Participating Institutions. The Participating Institutions are the American Museum of Natural History, Astrophysical Institute Potsdam, University of Basel, University of Cambridge, Case Western Reserve University, University of Chicago, Drexel University, Fermilab, the Institute for Advanced Study, the Japan Participation Group, Johns Hopkins University, the Joint Institute for Nuclear Astrophysics, the Kavli Institute for Particle Astrophysics and Cosmology, the Korean Scientist Group, the Chinese Academy of Sciences (LAMOST), Los Alamos National Laboratory, the Max-Planck-Institute for Astronomy (MPIA), the Max-Planck-Institute for Astrophysics (MPA), New Mexico State University, Ohio State University, University of Pittsburgh, University of Portsmouth, Princeton University, the United States Naval Observatory, and the University of Washington.

\end{document}